\documentclass[a4paper,fleqn,usenatbib]{mnras}

\usepackage{newtxtext,newtxmath}
\usepackage[T1]{fontenc}
\usepackage{ae,aecompl}

\usepackage{graphicx}	
\usepackage{amsmath}	

\title[Non-steady heating of cool cores]{Non-steady heating of cool cores of galaxy clusters by ubiquitous turbulence and AGN}

\author[Y. Fujita et al.]{
Yutaka Fujita,$^{1}$\thanks{Present address: Department of Physics, Graduate School of Science, 
Tokyo Metropolitan University,
1-1 Minami-Osawa, Hachioji-shi, Tokyo 192-0397, Japan, E-mail: y-fujita@tmu.ac.jp}
Renyue Cen,$^{2}$
and Irina Zhuravleva$^{3}$
\\
$^{1}$Department of Earth and Space Science, Graduate School of Science, Osaka University, Toyonaka, Osaka 560-0043, Japan\\
$^{2}$Department of Astrophysical Sciences, Princeton University, Princeton, NJ 08544, USA\\
$^{3}$Department of Astronomy \& Astrophysics, University of Chicago, Chicago, IL 60637, USA
}

\date{Accepted XXX. Received YYY; in original form ZZZ}

\pubyear{2019}

\begin{document}
\label{firstpage}
\pagerange{\pageref{firstpage}--\pageref{lastpage}}
\maketitle

\begin{abstract}
Recent cosmological simulations have shown that turbulence should be
generally prevailing in clusters because clusters are continuously
growing through matter accretion. Using one-dimensional
hydrodynamic simulations, we study the heating of cool-core clusters by
the ubiquitous turbulence as well as feedback from the central active
galactic nuclei (AGNs) for a wide range of cluster and
turbulence parameters, focusing on the global stability of the core. We
find that the AGN shows intermittent activities in the presence of
moderate turbulence similar to the one observed with
\textit{Hitomi}. The cluster core maintains a quasi-equilibrium state
for most of the time because the heating through turbulent diffusion is
nearly balanced with radiative cooling. The balance is gradually lost
 because of slight dominance of the radiative cooling, and the AGN is
ignited by increased gas inflow. Finally, when the AGN bursts, the core
is heated almost instantaneously. Thanks to the pre-existing turbulence,
the heated gas is distributed throughout the core without becoming globally unstable and causing catastrophic cooling,
and the core recovers the quasi-equilibrium state. The AGN bursts can be
stronger in lower-mass clusters.  Predictions of our model can be easily
checked with future X-ray missions like \textit{XRISM} and
\textit{Athena}.
\end{abstract}




\section{Introduction}
Radiative cooling time of the hot gas in intracluster medium (ICM) is
often shorter than the Hubble time in the central regions of many galaxy
clusters. In the absence of any heating sources, the hot gas in the core
should cool and a flow toward the cluster centre should develop (a
cooling flow; \citealt{1994ARA&A..32..277F}). However, X-ray
observations did not confirm the existence of massive cooling flows in
clusters, suggesting that the cores are heated by some unknown
sources
\citep[e.g.][]{1997ApJ...481..660I,2001A&A...365L.104P,2001A&A...365L..87T,2001A&A...365L..99K}. The
most promising candidate of the heating source is active galactic
nucleus (AGN) that resides in the cluster centre
\citep[e.g.][]{2000A&A...356..788C,2007ARA&A..45..117M,2012ARA&A..50..455F}. While
there seems to be consensus that AGNs provide enough energy to
counterbalance the radiative cooling, the question of how the energy is
conveyed to the surrounding ICM is still under debate. So far, various
mechanisms have been proposed, such as sound waves
\citep{2006MNRAS.366..417F,2017MNRAS.464L...1F,2018ApJ...858....5Z},
shocks \citep{2015ApJ...805..112R,2017ApJ...847..106L}, and cosmic-rays
\citep{1991ApJ...377..392L,2008MNRAS.384..251G,2012ApJ...746...53F,2013MNRAS.432.1434F,2013ApJ...779...10P,2017MNRAS.467.1449J,2017ApJ...844...13R,2020MNRAS.491.1190S}.
If a cool core is assumed to be in a steady state, global stability
needs to be considered. In fact, it is a serious problem for some
heating models. For example, non-linear evolution of large-amplitude
sound waves in clusters leads to localized heating, which destabilizes
the core \citep{2005ApJ...630L...1F,2006ApJ...638..659M}.

Turbulence may be another carrier of heat in the ICM
\citep{2003ApJ...596L.139K,2005ApJ...622..205D,2010ApJ...713.1332R,2011MNRAS.414.1493R,2014Natur.515...85Z}. Recently,
{\it Hitomi} has discovered moderate turbulence in the Perseus cluster
\citep{2016Natur.535..117H}. The ICM has a line-of-sight velocity
dispersion of $164\pm 10\rm\: km\: s^{-1}$ in the region $r=30$--60~kpc
from the cluster centre.  A similar level of turbulence in clusters has
been measured from fluctuations of the X-ray surface brightness
\citep{2004A&A...426..387S,2014Natur.515...85Z,2018ApJ...865...53Z}. Since
{\it Hitomi} observed only a few small regions inside the core, it is
unclear whether the turbulence is created through the activities of the
central AGN or not\footnote{Recent results from optical observations
indicated that turbulence at the centres of three clusters is driven by
the AGNs \citep{2020ApJ...889L...1L}}.

Here, we point out that AGNs are likely not the only source of
turbulence in and around cluster cool cores. Cosmological numerical
simulations have shown that the level of turbulence discovered by {\it
Hitomi} can be explained by cluster formation
\citep{2017ApJ...849...54L,2018PASJ...70...51O}. This reflects the fact
that clusters are still growing, and gas and dark matter (as galaxies)
are intermittently falling into them. This causes variations in cluster
potential resulting in gas sloshing
\citep{2001ApJ...562L.153M,2006ApJ...650..102A,2019ApJ...882..119Z}, and
excites turbulence in the ICM
\citep{2004ApJ...612L...9F,2005ApJ...619L.139F,2013ApJ...762...78Z} even
in apparently relaxed clusters
\citep{2017ApJ...849...54L,2018PASJ...70...51O,2020ApJ...892..100U}. This
ubiquitous turbulence is expected to convey thermal energy not only from
the central AGN but also from the outside to the inside of a
core. However, fairly strong turbulence (with the Mach number of
$\gtrsim 0.3$ on scales $\gtrsim 100$~kpc;
\citealp{2014Natur.515...85Z}) may be required to counterbalance the
radiative cooling if AGN feedback is ignored.

In this study, we consider a heating model that incorporates the effects
of simultaneous heating by both the central AGN and moderate turbulence
excited by matter accretion onto clusters using one-dimensional (1D)
numerical simulations. Using three-dimensional (3D)
simulations, \citet{2017ApJ...849...54L} and \citet{2017MNRAS.472.4707B}
studied these two processes and indicated that large-scale bulk
and shear motions associated with the growth of clusters can
enhance the mixing and advection of AGN feedback energy. 
Instead of performing computationally-expensive 3D simulations, we will
use 1D models to explore a wide range of cluster and turbulence
parameters in order to find general trends of AGN feedback behaviour in
turbulent ICM.
We focus on {\it global} core-scale
stability and we do {\it not} assume that a cool core is in a steady
state. This approach is different from that taken in some
multi-dimensional simulations in which the heating and cooling is
assumed to be globally balanced in order to focus on {\it local} thermal
instability \citep[e.g.][]{2012MNRAS.419.3319M,2012MNRAS.420.3174S}.
Our model is similar to a model that combines AGN feedback and thermal
conduction \citep{2002ApJ...581..223R}.  However, contrary to the
latter, our model shows that a cluster core does not approach a steady
state, and the central AGN intermittently bursts.

This paper is organized as follows. In Section 2, we describe our models
on turbulence and AGN feedback. In Section 3, we show that the results
of our numerical simulations. In Section 4, we discuss the implications
of our results. Finally, Section 5 is devoted to conclusions. We assume
a $\Lambda$CDM cosmology with $H_0 = 70 \rm\: km s^{-1} Mpc^{-1}$
($h=0.7$), $\Omega_0 = 0.3$ and $\lambda=0.7$.

\section{Models}

\subsection{Hydrodynamic equations}

We assume that clusters are spherically symmetric for the sake of
simplicity. The flow equations are
\begin{equation}
\label{eq:cont}
 \frac{\partial \rho}{\partial t} 
+ \frac{1}{r^2}\frac{\partial}{\partial r}(r^2\rho v) = \dot{\rho_*}\:,
\end{equation}
\begin{equation}
\label{eq:mom}
\frac{\partial (\rho v)}{\partial t} 
+ \frac{1}{r^2}\frac{\partial}{\partial r}(r^2\rho v^2)
= - \rho g-\frac{\partial P}{\partial r} - \dot{\rho_*}v\:,
\end{equation}
\begin{eqnarray}
\label{eq:eg}
 \frac{\partial e_g}{\partial t}  
+ \frac{1}{r^2}\frac{\partial}{\partial r}(r^2 v e_g)
&=& -P \frac{1}{r^2}\frac{\partial}{\partial r}(r^2 v) 
+ \frac{1}{r^2}\frac{\partial}{\partial r}
\left(r^2\kappa\frac{\partial T}{\partial r}\right)\nonumber\\
&+& \frac{c_{\rm diss}\rho u^3}{l} 
+ \frac{1}{r^2}\frac{\partial}{\partial r}
\left(r^2 D_{\rm eddy}\rho T\frac{\partial s}
{\partial r}\right)
\nonumber\\
&-& n_e^2\Lambda(T) + h_{\rm AGN} - c_*\:,
\end{eqnarray}
where $t$ is the time, $r$ is the cluster centric radius, $\rho$ is the
gas density, $v$ is the bulk velocity, $P$ is the pressure, $T$ is the
temperature, and $s$ is the specific entropy. For other parameters,
$\dot{\rho_*}$ is the mass-loss rate of stars per unit volume, $u$ is
the turbulent velocity, $l$ is the dominant scale of turbulence, $g$ is
the gravitational acceleration, $\kappa$ is the thermal conductivity,
$D_{\rm eddy}$ is the the eddy diffusivity, $c_{\rm diss}$ is a
dimensionless constant, $n_e$ is the electron number density, and
$\Lambda$ is the cooling function, $h_{\rm AGN}$ is the heating by the
AGN, and $c_*$ is the cooling due to mass-loss of stars in the brightest
cluster galaxy (BCG) at the cluster centre. The energy density of gas is
defined as $e_g=P/(\gamma-1)$, where $\gamma=5/3$. The second, third,
and forth terms of the right hand side of equation~(\ref{eq:eg})
represent thermal conduction, turbulent dissipation, and turbulent
diffusion, respectively.  While the thermal energy of the ICM
is carried by electrons for thermal conduction, it is conveyed by
turbulent eddies for turbulent diffusion. The turbulent diffusion mimics
the mixing of gas in hotter regions with that in cooler regions. The
turbulent dissipation means viscous dissipation of turbulent
motion. Since these terms are clearly separated in
equation~(\ref{eq:eg}), we can easily estimate their relative
contribution to the heating, which is not so simple for
multi-dimensional simulations. The thermal conductivity is given by
\begin{equation}
\label{eq:kappa}
 \kappa(T) = 5\times 10^{-7}f_c(T/{\rm K})^{5/2}
\end{equation}
in cgs units, where $f_c$ is a reduction factor compared with the
Spitzer value.

We assume that turbulence is induced through the growth of clusters and
thus the turbulent velocity $u$ should be related to the depth of the
gravitational potential well. For simplicity, we assume that the AGN
feedback does not affect the velocity $u$.
Thus, we assume that
\begin{equation}
\label{eq:u}
 u(r) = \alpha_u V_{\rm cir}(r)\:,
\end{equation}
where $\alpha_u$ is a parameter and $V_{\rm cir}$ is a circular velocity
defined by the gravitational potential of the cluster (see
equation~[\ref{eq:ccir}]). 
This gives an outwardly increasing velocity profile in the region we are interested in (see Figure~\ref{fig:u1D}), which is consistent with the simulation results of \citet[their Figure~8]{2018PASJ...70...51O}
For turbulent dissipation, we assume that
$c_{\rm diss}=0.42$ \citep{2005ApJ...622..205D} and 
\begin{equation}
\label{eq:l}
l = \alpha_l r\:,
\end{equation}
where $\alpha_l$ is an adjustable constant following
\citet{2003ApJ...596L.139K} and \citet{2005ApJ...622..205D}.

The eddy diffusivity is given by
\begin{equation}
 D_{\rm eddy} = c_{\rm td}u l \xi\:,
\end{equation}
where $c_{\rm td}$ is a dimensionless constant and $\xi$ is a
reduction factor. Following \citet{2005ApJ...622..205D}, we adopt
$c_{\rm td}=0.11$ (see also \citealp{1994PhFl....6.3416Y}). The
reduction factor is introduced considering the effects of buoyancy. If
the dominant eddy-turnover frequency $u/l$ is much smaller than the
Brunt-V\"ais\"al\"a frequency,
\begin{equation}
 N_{\rm BV} = \sqrt{g\left(\frac{1}{\gamma P}\frac{d P}{d r}
-\frac{1}{\rho}\frac{d\rho}{d r}
\right)}\:,
\end{equation}
then the radial motions are primarily determined by buoyancy
oscillations with frequency $N_{\rm BV}$. In this case, the radial
displacement of a fluid element is $\sim u/N_{\rm BV}$, which is much
smaller than the eddy size $l$. Thus, the radial diffusion of heat is
significantly suppressed by a factor of
\begin{equation}
 \label{eq:reduc}
\xi = \frac{1}{1+c_0^2 l^2 N_{\rm BV}^2/u^2}\:,
\end{equation}
where $c_0^2 = 0.1688$ \citep{1981JGR....86.9925W,2005ApJ...622..205D}.

For the cooling function, we adopt the following metallicity-dependent
function:
\begin{eqnarray}
\Lambda(T,Z)&=&2.41\times 10^{-27}
\left[0.8+0.1\left(\frac{Z}{Z_\odot}\right)\right]
\left(\frac{T}{\rm K}\right)^{0.5}\nonumber\\
&+ & 1.39\times 10^{-16}
\left[0.02+0.1\left(\frac{Z}{Z_\odot}\right)^{0.8}\right]\nonumber\\
& & \times\left(\frac{T}{\rm K}\right)^{-1.0}\rm\: erg\: cm^{3}\:
\label{eq:cool}
\end{eqnarray}
\citep{2013MNRAS.428..599F}, which approximates the cooling function
derived by \citet{1993ApJS...88..253S} for $T\ga 10^5$~K and $Z\la 1\:
Z_\odot$. We assume that equation~(\ref{eq:cool}) can be applied even for $T\la 10^5$~K. In our calculations, however, the temperature of gas does not fall to $10^5$~K, except one model (M25, see Section~\ref{sec:para-dep}).
Since we are interested in the central region of clusters, we
adopt $Z=0.5\rm Z_\odot$, hereafter.

\subsection{Cluster potential}
\label{sec:pot}

We assume that the potential of clusters is given by the NFW profile
\citep{1997ApJ...490..493N}. The density profile is given by
\begin{equation}
\label{eq:NFW}
 \rho_{\rm DM}(r) = \frac{\delta_c\rho_c}{(r/r_s)(1+r/r_s)^2}\:,
\end{equation}
where $r_s$ is the characteristic radius, $\rho_c$ is the critical
density of the Universe, and $\delta_c$ is the normalization. We define
the characteristic mass $M_s$ as the mass enclosed within $r=r_s$.  The
halo concentration parameter is given by
\begin{equation}
 \label{eq:cD}
c_\Delta=r_\Delta/r_s\:,
\end{equation}
where $r_\Delta$ is the radius inside which the average density is
$\Delta$ times the critical density of the Universe $\rho_c(z)$. 
From this definition, the mass inside $r_\Delta$ is given by
\begin{equation}
\label{eq:MD}
 M_\Delta = \frac{4\pi}{3}\Delta \rho_c(z)r_\Delta^3\:.
\end{equation}
The mass profile of the NFW profile is written as
\begin{equation}
\label{eq:MNFW}
 M_{\rm NFW}(r) = 4\pi\delta_c\rho_c r_s^3
\left[\ln\left(1+\frac{r}{r_s}\right)-\frac{r}{r+r_s}\right]\:.
\end{equation}
From this equation, the characteristic mass $M_s$ can be expressed in
terms of $M_\Delta$ and $c_\Delta$:
\begin{equation}
\label{eq:MDMs}
 M_s = M_\Delta\frac{\ln 2-1/2}{\ln(1+c_\Delta)-c_\Delta/(1+c_\Delta)}\:.
\end{equation}

For a given $M_\Delta$, the characteristic radius $r_s$, the mass $M_s$
inside $r_s$, and the typical X-ray temperature in the inner region of
the cluster $T_c$ can be determined as follows. From the CLASH massive
cluster sample \citep{2012ApJS..199...25P}, \citet{2018ApJ...857..118F}
found that the cluster X-ray temperature $T_c$ has a tight correlation
with $r_s$ and $M_s$:
\begin{equation}
\label{eq:TX}
 T_c = T_{c0}\left(\frac{r_s}{r_{s0}}\right)^{-2}
\left(\frac{M_s}{M_{s0}}\right)^{3/2}\:,
\end{equation}
where $(r_{s0}, M_{s0}, T_{c0})$ is a representative point of the
relation \citep{2018ApJ...863...37F}.\footnote{We use $r_{s0}=414$~kpc,
$M_{s0}=1.4\times 10^{14}\: M_\odot$, and $T_{c0}=3.7$~keV based
on the results of the MUSIC simulations
\citep{2014ApJ...797...34M,2018ApJ...857..118F}.}  Note that $T_c$ is
the temperature at $r=50$--500~kpc from the cluster centre. For
$\Delta=200$, the concentration parameter is represented by
\begin{equation}
\label{eq:duf08a}
 c_{200}(M_{200},z) = 6.71\:\left(\frac{M_{200}}
{2\times 10^{12}h^{-1}M_\odot}\right)^{-0.091}(1+z)^{-0.44}
\end{equation}
for $M_{200}\sim 10^{11}$--$10^{15}h^{-1}M_\odot$ and $z<2$
(\citealp{2008MNRAS.390L..64D}; see also
\citealp{2013ApJ...766...32B,2014MNRAS.441.3359D,2014ApJ...797...34M,2015ApJ...799..108D,2015MNRAS.452.1217C}).
This relation has a large dispersion, which makes a variety of $T_c$ for
a given mass \citep{2018ApJ...863...37F} but we ignore it here. For a
given $z$ and $M_{200}$, $r_s$ and $M_s$ can be derived from
equations~(\ref{eq:cD}), (\ref{eq:MD}), (\ref{eq:MDMs}) and (\ref{eq:duf08a}), and thus
$T_c$ is obtained from equation~(\ref{eq:TX}). We assume $z=0$,
hereafter.

We also include the stellar contribution of the brightest cluster galaxy
(BCG) to the gravitational acceleration. For the Perseus cluster
($M_{200}\sim 8.5\times 10^{14}\: M_\odot$;
\citealp{2006ApJ...638..659M}), it is
\begin{equation}
 g_{*0}(r) = \left[\left(\frac{r^{0.5975}}{3.206\times 10^{-7}}\right)^{9/10}
+ \left(\frac{r^{1.849}}{1.861\times 10^{-6}}\right)^{9/10}
\right]^{-10/9}\:\rm cm\; s^{-2}\:,
\end{equation}
where $r$ is in kpc \citep{2006ApJ...638..659M}. Since the mass of a BCG
is weakly correlated to the mass of the host halo ($M_{\rm BCG}\propto
M_{200}^{0.4}$; \citealp{2018AstL...44....8K,2019A&A...631A.175E}), we
scale the acceleration as $g_* = g_{*0}(M_{200}/8.5\times 10^{14}\:
M_\odot)^{0.4}$. Thus, the total acceleration is
\begin{equation}
 \label{eq:gtot}
g(r) = g_*(r) + \frac{G M_{\rm NFW}(r)}{r^2}\:,
\end{equation}
where $G$ is the gravitational constant. The stellar mass profile of the
BCG is written as 
\begin{equation}
\label{eq;Mst}
 M_*(r)=g_*(r)r^2/G\:.
\end{equation}
The circular velocity of the cluster is given by
\begin{equation}
\label{eq:ccir}
 V_{\rm cir}(r) = \sqrt{\frac{GM(r)}{r}} = \sqrt{g(r) r}\:.
\end{equation}
where $M(r)=M_*(r)+M_{\rm NFW}(r)$ is the total mass profile. The
velocity profile is time-independent.

\subsection{AGN heating}
\label{sec:AGN}

We adopt a simple model for AGN heating.  The injection of the
energy is to mimic the heating of the ICM by a central source and not to
model a specific mechanism of AGN feedback such as that through the hot
jet lobes.  We do not specify the main carrier of the energy injected by
the AGN. For example, if AGN jets are the energy source, gas motion is
likely to be created. On the other hand, if cosmic-rays accelerated in
the vicinity of the central black hole are the carrier, 
they will stream with Alfv\'en waves in the ICM 
which will be damped into heat
\citep{2008MNRAS.384..251G,2013MNRAS.432.1434F}\footnote{These studies indicate that the heating by cosmic-rays is fairly stable without turbulence. We mean that turbulence further contributes to the stability.}. If sound waves are the
carrier, the ICM will be heated through the viscous dissipation; the ICM
may just oscillate and strong turbulence may not be
generated.  In this study, we do not explicitly 
simulate the possible gas
motion created by the AGN activities for the sake of simplicity.

We assume that the total heating rate depends on the mass accretion rate
$\dot{M}=-4\pi r^2\rho v$ ($v<0$) at the inner boundary of the
simulation region ($r=r_{\rm in}$). The Eddington accretion rate
when a radiative efficiency is unity is given by
\begin{equation}
\label{eq:dotMEdd}
 \dot{M}_{\rm Edd} = L_{\rm Edd}/c^2\:,
\end{equation}
where $L_{\rm Edd}=1.3\times 10^{38}(M_{\rm BH}/M_\odot)\rm\: erg\:
s^{-1}$ is the Eddington luminosity and $M_{\rm BH}$ is the black hole
mass\footnote{In \citet{2014ARA&A..52..529Y}, the Eddington
accretion rate is defined as $\dot{M}_{\rm Edd} =10\: L_{\rm Edd}/c^2$,
which means that a radiative efficiency of 0.1 is assumed. For this
definition, an advection-dominated accretion flow is realized at
$\dot{M}/\dot{M}_{\rm Edd} \lesssim 0.01$.}.  We assume that the black
hole mass is proportional to the stellar mass of the BCG. Since the mass
of the black hole at the centre of the Perseus cluster is $1\times
10^9\: M_\odot$ \citep{2019ApJ...883..193N}, we assume that $M_{\rm
BH}=1\times 10^9\: M_\odot (M_{200}/8.5\times 10^{14}\:
M_\odot)^{0.4}$. The total ICM heating rate is given by
\begin{equation}
\label{eq:LAGN}
\dot{E}_{\rm ICM} = \epsilon\eta(\dot{M}) \dot{M}c^2\:,
\end{equation}
where $\epsilon<1$ is the heating efficiency and $\eta(\dot{M})$ is a
correction factor. We introduce $\epsilon$ because not all of the
rest-mass energy of the gas engulfed by the black hole is used to heat
the ICM.  Moreover, not all the gas that passed the inner boundary
($r_{\rm in}$) reaches the black hole; some of it will form stars in the
BCG or may 
never reach or become bound
to the black hole. The correction factor reflects the fact that the
nature of an accretion flow changes at $\dot{M}/\dot{M}_{\rm Edd} \sim
0.1$ \citep{2014ARA&A..52..529Y}. In particular, an advection-dominated
accretion flow that is realized at $\dot{M}/\dot{M}_{\rm Edd} \lesssim
0.1$ has a low radiative efficiency. Thus, we assume that $\eta = 1$ for
$\dot{M}/\dot{M}_{\rm Edd}>0.1$ and
\begin{equation}
\label{eq:eta}
 \eta = \frac{\dot{M}}{0.1\dot{M}_{\rm Edd}}
\end{equation}
for $\dot{M}/\dot{M}_{\rm Edd}<0.1$ following
\citet{2014ARA&A..52..529Y}. Even if we assume that $\eta = 1$
regardless of $\dot{M}$, results do not change much. We also note
that even if all the inflow gas $\dot{M}$ is added to $M_{\rm BH}$
during the calculation period (8~Gyr; see section~\ref{sec:result}), the
total mass is comparable to $M_{\rm BH}$ and the results are almost the
same. 

The heating rate per the unit volume is given by
\begin{equation}
 h_{\rm AGN}(r) =  
\frac{\dot{E}_{\rm ICM}}{4\pi (r_{\rm AGN}-r_{\rm in}) r^2}\:,
\end{equation}
where $r_{\rm AGN}$ is the maximum radius inside which the AGN heating
is effective. For $r>r_{\rm AGN}$, we assume $h_{\rm AGN}(r)=0$. The
radial dependence of $h_{\rm AGN}\propto r^{-2}$ may be realized if the
ICM is heated at the surface of some kind of waves or fronts of which
surface area increases as $\propto r^2$. The radius of the heated region
is assume to be $r_{\rm AGN}=50$~kpc, based on the fact that the central
AGN disturbs the ambient ICM on a scale of $\sim 50$~kpc for the Perseus
cluster \citep{2007MNRAS.381.1381S}. Note that if the heating rate is
represented by $h_{\rm AGN}\propto r^{-\beta}$, $\beta$ and $r_{\rm
AGN}$ are degenerated. If $\beta (>0)$ is larger and/or $r_{\rm AGN}$ is
smaller, the heating is more centrally concentrated. We find that if the
heating is too much concentrated (e.g. $\beta=2$ and $r_{\rm
AGN}$=20~kpc), the turbulent mixing of hot and cool gas is not enough and the stability of the ICM is lost.

\subsection{Stellar mass-loss}
\label{sec:ML}

The gas ejected from stars in the BCG mixes with the surrounding ICM.
This serves as cooling in equation~(\ref{eq:eg}), because the mass-loss
gas is cooler than the ambient ICM. This gas may be the cause of
multi-temperature structure of the ICM around the cluster centre.

The cooling rate due to the mass-loss gas is given by
\begin{equation}
 c_* = -\alpha_*\rho_*\left(\epsilon_0 - \frac{P}{\rho} - \epsilon_{\rm ICM} 
+ \frac{1}{2}v^2\right)\:,
\end{equation}
where $\epsilon_{\rm ICM}=3kT/(2 \mu m_p)$ is the specific thermal
energy of the ICM, $k$ is the Boltzmann constant, and $\mu = 0.61$ is
the mean molecular weight \citep{2003ARA&A..41..191M}. The specific mass
loss-rate is $\alpha_*=4.7\times 10^{-20}\rm\: s^{-1}$ at $z\sim 0$
\citep{2003ARA&A..41..191M}. The stellar mass density $\rho_*(r)$ can be
derived from equation~(\ref{eq;Mst}) and the mass-loss rate is written
as $\dot{\rho}_* = \alpha_*\rho_*$ in equation~(\ref{eq:mom}).  The
source term $\alpha_*\rho_*(\epsilon_0 -P/\rho - \epsilon_{\rm ICM})$
represents the heating of the hot ICM of specific energy $\epsilon_{\rm
ICM}$ by the mean energy of stellar ejecta $\epsilon_0$ less the work
done $P/\rho$ in displacing the hot gas. The term $\alpha_*\rho_* v^2/2$
represents dissipative heating. The mean gas injection energy is
$\epsilon_0 = 3kT_0/(2 \mu m_p)$, where $T_0=(\alpha_* T_* + \alpha_{\rm
sn} T_{\rm sn})/\alpha_*$. The stellar temperature mainly reflects the
kinetic energy of stars and is given by $T_*=(\mu m_p/k)\sigma_*^2$,
where $\sigma_*$ is the stellar velocity dispersion of the BCG. Here, we
assume a typical value of $\sigma_*=275\:\rm km\: s^{-1}$
\citep{2009MNRAS.398..133L} and the following results are not sensitive
to the value. The characteristic temperature of supernovae is written as
$T_{\rm sn}=2 \mu m_p E_{\rm sn}/(3 k M_{\rm sn})$, where $E_{\rm
sn}=10^{51}\rm\: erg$ is the explosion energy of a supernova and $M_{\rm
sn}$ is the ejecta mass. The specific mass loss rate from supernovae is
$\alpha_{\rm sn}\approx 2\times 10^{-22}(M_{\rm sn}/M_\odot)\rm\:
s^{-1}$ at $z\sim 0$, in which the mass-to-light ratio is assumed to be
$\sim 10$ \citep{2003ARA&A..41..191M}. The product $\alpha_{\rm sn}
T_{\rm sn}$ does not depend on $M_{\rm sn}$ and $\alpha_{\rm sn}$ is
much smaller than $\alpha_*$. If the cluster mass is
$M_{200}\neq 8.5\times 10^{14}\: M_\odot$, the stellar mass density is
multiplied by a factor of $(M_{200}/8.5\times 10^{14}\:
M_\odot)^{0.4}$.

\begin{table}
\centering
\caption{Model parameters}
\label{tab:model}
\begin{tabular}[t]{lcccccc}
		\hline
Model & $M_{200}$ & $\alpha_u$ & $\alpha_l$ & variation & $\epsilon$ & $f_c$ \\
 & ($10^{14}\: M_\odot$) &  &  &  &  &  \\
		\hline
M35 & 8.5 & 0.3 & 0.5 & no & 0.01 & 0 \\
M25 & 8.5 & 0.2 & 0.5 & no & 0.01 & 0 \\
M55 & 8.5 & 0.5 & 0.5 & no & 0.01 & 0 \\
M33 & 8.5 & 0.3 & 0.3 & no & 0.01 & 0 \\
M35v & 8.5 & 0.3 & 0.5 & yes & 0.01 & 0 \\
M35e & 8.5 & 0.3 & 0.5 & no & 0.1 & 0 \\
M25c & 8.5 & 0.2 & 0.5 & no & 0.01 & 0.2 \\
L35/L35b & 5.5 & 0.3 & 0.5 & no & 0.01 & 0 \\
H35 & 14 & 0.3 & 0.5 & no & 0.01 & 0 \\
		\hline
\end{tabular}
\end{table}

\section{Results}
\label{sec:result}

The hydrodynamic equations (\ref{eq:cont})-(\ref{eq:eg}) are solved by a
second-order advection upstream splitting method (AUSM) based on
\citeauthor{1993JCoPh.107...23L} (1993; see also
\citealt{2001ApJ...547..172W}).  We set the inner and outer boundary at
$r_{\rm min}=3$ and $r_{\rm max}=500$~kpc, respectively. 
Since the innermost regions are often strongly asymmetrically disturbed by the AGN activity \citep[e.g.][]{2001ApJ...558L..15B}, we cannot reproduce them in our 1D simulations. Therefore, we do not consider the innermost $<3$~kpc regions.  We use 300
unequally spaced meshes in the radial coordinate to cover the region.
The innermost mesh has a width of $\sim 40$~pc, and the width of the
outermost mesh is $\sim 10$~kpc. The following boundary conditions are
adopted. (1) Variables except velocity have zero gradients at the inner
boundary. (2) An outflow boundary is adopted for the velocity
at the inner boundary. (3) The density and pressure at the outermost
mesh are equal to specified initial values. Initially, the ICM is in a
hydrostatic equilibrium and isothermal at $T=T_c$ and the gas
mass fraction for $r<r_{\rm max}$ is 0.13, which is consistent with
observations
\citep[e.g.][]{2006A&A...452...75B,2013ApJ...778...14G,2015MNRAS.450..896D}.

We solve the hydrodynamic equations without turbulence and thermal conduction until temperature at the inner boundary drops to
$T=2$~keV in order to create a cool core quickly. Then, we turn on
turbulence and thermal conduction, and set this time for $t=0$. We
calculate until $t=8$~Gyr, which is the look-back time of $z\sim 1$.

\begin{figure}
 \includegraphics[width=0.9\columnwidth]{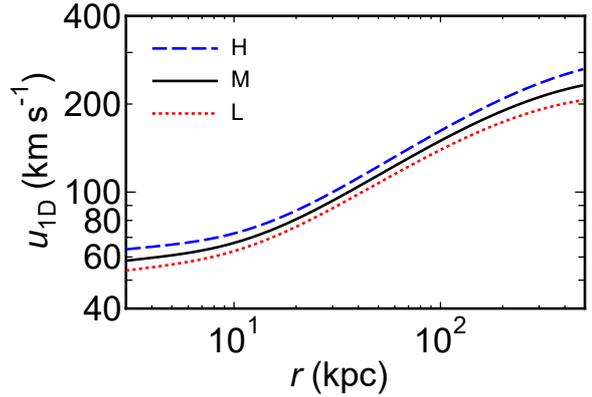} \caption{
 Mass-dependence of the profiles for the 1D turbulent velocity when
 $\alpha_u=0.3$. Blue dashed, solid black, and dotted red lines show
 the cases of $M_{200}=1.4\times 10^{15}$, $8.5\times 10^{14}$, and
 $5.5\times 10^{14}\: M_\odot$, respectively.}  \label{fig:u1D}
\end{figure}

\begin{figure}
 \includegraphics[width=0.9\columnwidth]{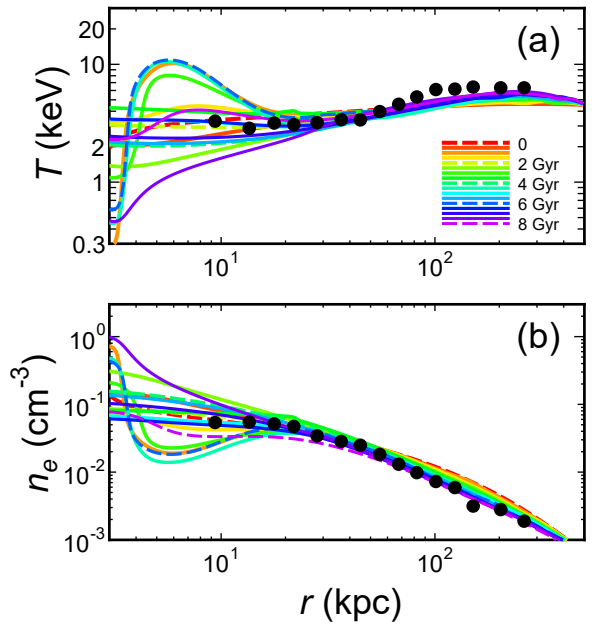} \caption{
    Evolution of the (a) temperature and (b) electron density profiles
    for the fiducial model M35 -- sampled every 0.5~Gyr, from red to
    purple.  Dashed lines correspond to specific times ($t=0$, 2, 4, 6,
    and 8~Gyr).  Black dots are observations of the Perseus cluster
    \citep{2015MNRAS.450.4184Z}. The errors are smaller than the size of the dots.}  \label{fig:Tn_M35}
\end{figure}

\begin{figure}
 \includegraphics[width=0.9\columnwidth]{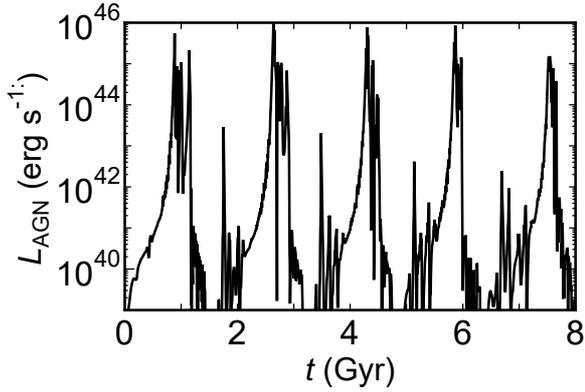} \caption{Variation of
    AGN power for the fiducial model M35.}
    \label{fig:LAGN_M35}
\end{figure}

\begin{figure}
 \includegraphics[width=0.9\columnwidth]{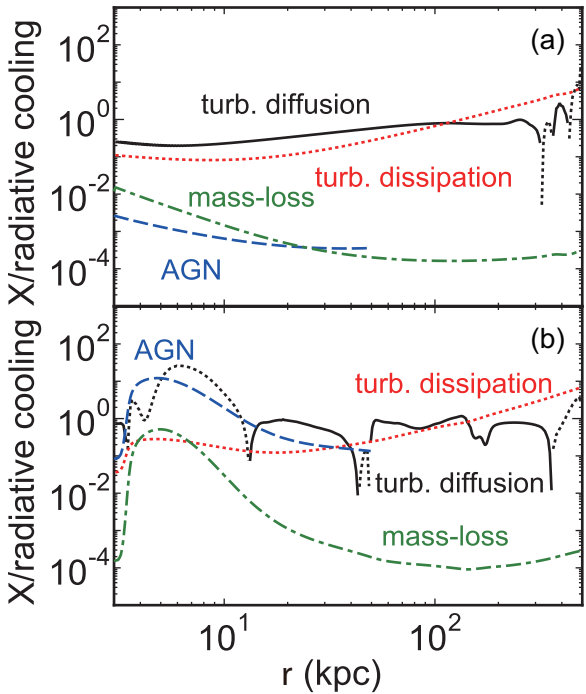} \caption{Ratios of
 heating and stellar mass-loss terms ({\it X}) to the radiative cooling
 term ($n_e^2\Lambda(T)$) on the right hand side of
 equation~(\ref{eq:eg}) at (a) $t=4$~Gyr and (b) $t=1$~Gyr for 
 the fiducial model M35. Dashed red line: turbulent dissipation (third
 term).  Solid black line: turbulent diffusion when the fourth
 term is positive (heating). Dotted black line: turbulent diffusion when
 the fourth term is negative (cooling).  Dashed blue line: AGN heating
 ($h_{\rm AGN}$). Dash-dotted green line: stellar mass-loss ($c_*$).}
 \label{fig:balance_M35}
\end{figure}

\subsection{Fiducial model}

Table~\ref{tab:model} shows our models and parameters.  As a fiducial
model (M35), we choose turbulence parameters as $\alpha_u=0.3$
(equation~[\ref{eq:u}]) and $\alpha_l=0.5$ (equation~[\ref{eq:l}]). The
cluster mass is $M_{200}=8.5\times 10^{14}\: M_\odot$, which gives
$c_{200}=4.0$ and $T_c=4.6$~keV (section~\ref{sec:pot}). The
accretion efficiency for the AGN heating is $\epsilon=0.01$
(equation~[\ref{eq:LAGN}]) and thermal conduction is ignored ($f_c=0$;
equation~[\ref{eq:kappa}]).

Figure~\ref{fig:u1D} shows the radial profile of the 1D turbulent
velocity $u_{\rm 1D} \equiv u/\sqrt{3}$ given by
equation~(\ref{eq:u}). The velocity is $u_{\rm 1D}\sim 110\rm\: km\:
s^{-1}$ at $r\sim 50$~kpc, which is comparable to the value derived from
the latest analysis of the {\it Hitomi} data for the Perseus cluster
\citep{2018PASJ...70....9H}. The velocity gradually increases outward as
is predicted by cosmological numerical simulations
\citep[e.g.][]{2018PASJ...70...51O}. {\it Hitomi} observations have also
shown that the spatial scale of turbulence is $<100$~kpc for $r\lesssim
100$~kpc \citep{2018PASJ...70....9H}. Thus, our choice of $\alpha_l=0.5$
is consistent with the result. Our assumption is also consistent with
the amplitude of motions estimated from density fluctuations in the
central regions of ten clusters ($\sim 100$--$300\rm\: km\: s^{-1}$ on
scales of $\lesssim 100$~kpc; \citealp{2018ApJ...865...53Z}).

Figure~\ref{fig:Tn_M35} shows the profiles of temperature and density at
various times. The temperature and density at $r\lesssim 10$~kpc
fluctuate wildly owing to the changing AGN activities shown in
Figure~\ref{fig:LAGN_M35}. In Figure~\ref{fig:Tn_M35}, we plotted
observational data for the Perseus cluster \citep{2015MNRAS.450.4184Z}
as a reference. Although we do not fine-tune the cluster parameters
(e.g. $c_{200}$), our results broadly reproduce the
observations. Figure~\ref{fig:Tn_M35} indicates that the temperature
sometimes reaches $\sim 10$~keV at $r\lesssim 10$~kpc at strong bursts
of the AGN ($t\sim 1.0$, 2.6, 4.3, 5.9, and 7.5~Gyr in
Figure~\ref{fig:LAGN_M35}). During the bursts, the cool core could be
significantly disturbed, although there is a caveat that recent high-resolution 3D simulations find that cool cores
can survive AGN feedback episodes \citep[e.g.][]{2019MNRAS.490..343B,2020arXiv200106532C}.
Turbulent diffusion gradually smooths out the hot
region and distributes the
energy injected by the bursts across the core.  This process is similar
to a previous mixing model
\citep{2016MNRAS.455.2139H,2017MNRAS.466L..39H,2019arXiv191204349H}. In
their model, however, gas is mixed through gas motion generated by AGN
activities contrary to our model where turbulent mixing is driven by
structure formation. We note that our 1D results shown in
Figure~\ref{fig:Tn_M35} implicitly assume that gas is well-mixed in the
tangential direction and could not directly be compared with
observations for the innermost region of clusters ($\lesssim 10$~kpc),
where multi-temperature structures and bubbles have been observed. In
that region, hot tenuous gas and cool dense gas often coexist at the
same radius \citep[e.g.][]{2004MNRAS.349..952S}, and the azimuthally
averaged temperature and density are observationally biased toward the
cool dense gas. This means that while we simulate global
evolution of the core, observations could be biased by local
structures.  The temperature of the gas inside of the bubbles
may be much higher than 10~keV \citep{2019ApJ...871..195A}.

The periodical activities of the AGN shown in Figure~\ref{fig:LAGN_M35}
indicate that radiative cooling is not balanced with the AGN heating at
a given time. The excessive cooling increases $\dot{M}$ and $\dot{E}_{\rm
ICM}$ (equation~[\ref{eq:LAGN}]). Figure~\ref{fig:balance_M35}a shows
the ratios of heating terms to the radiative cooling term
$n_e^2\Lambda(T)$ in equation~(\ref{eq:eg}) at $t=4$~Gyr when the AGN
activity is not strong (Figure~\ref{fig:LAGN_M35}). The turbulent
diffusion term is the most important heating source for $r\lesssim
100$~kpc and the turbulent dissipation term comes next.  The
turbulent diffusion carries the heat from the outside to the inside of
the core.  Contributions of the AGN and stellar
mass-loss are negligible. If the two turbulence terms are combined, the
ratio to the radiative cooling becomes close to one but still smaller
than unity. This means that the cluster core is in a quasi-equilibrium
state. However, since the turbulent heating (dissipation plus diffusion)
does not fully counterbalance the radiative cooling, the cooling
gradually overwhelms the heating and the mass inflow rate $\dot{M}$
increases. This finally leads to strong AGN
bursts. Figure~\ref{fig:balance_M35}b shows the ratios at $t=1$~Gyr when
the AGN is active (Figure~\ref{fig:LAGN_M35}). The AGN heating exceeds
radiative cooling at $r\lesssim 10$~kpc. In this region, the
turbulent diffusion transports the heat outward
and, therefore, serves a source of cooling in this innermost region.

In Figure~\ref{fig:balance_M35}, turbulent dissipation surpasses
turbulent diffusion at $r\gtrsim 100$~kpc. This leads to gradual
increase of temperature in that region (Figure~\ref{fig:Tn_M35}). In a
real cluster this can be attributed to the increase of mass and
temperature of the cluster because the turbulence is induced by matter
accretion from the outside of the cluster. It is well known that when a
cluster forms, the kinetic energy associated with the bulk motion of
infalling gas is converted to thermal energy at shocks. The turbulent
heating is another process of the energy conversion. We expect that the
turbulent energy dissipation rate is related to the mass accretion rate
of the cluster.

\begin{figure}
 \includegraphics[width=0.9\columnwidth]{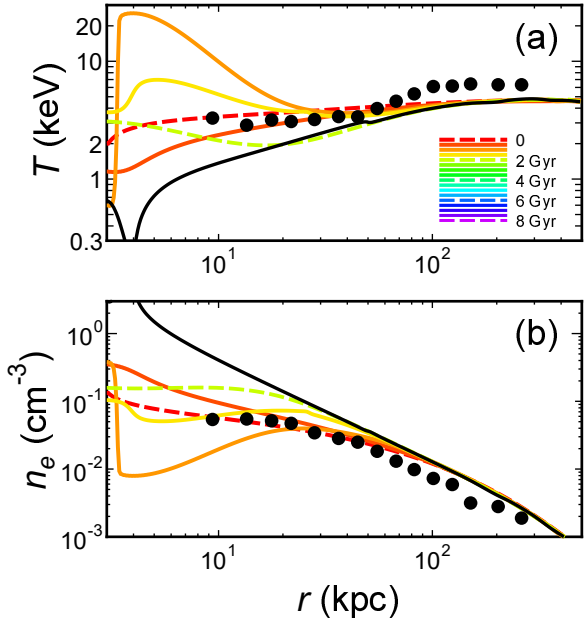} \caption{Same as
 Figure~\ref{fig:Tn_M35} but for model~M25 (smaller turbulent
 velocity). Solid black line shows $t=2.2$~Gyr or the end of the
 calculation.}  \label{fig:Tn_M25}
\end{figure}

\begin{figure}
 \includegraphics[width=0.9\columnwidth]{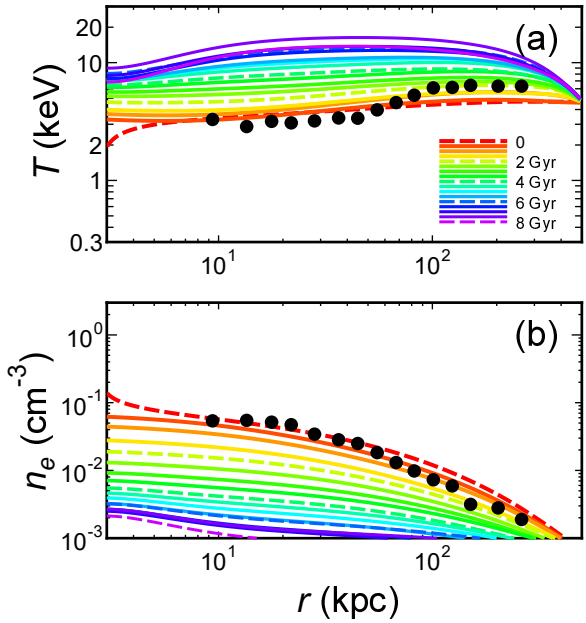} \caption{Same as
 Figure~\ref{fig:Tn_M35} but for model~M55 (larger turbulent
 velocity).}  \label{fig:Tn_M55}
\end{figure}

\begin{figure}
 \includegraphics[width=0.9\columnwidth]{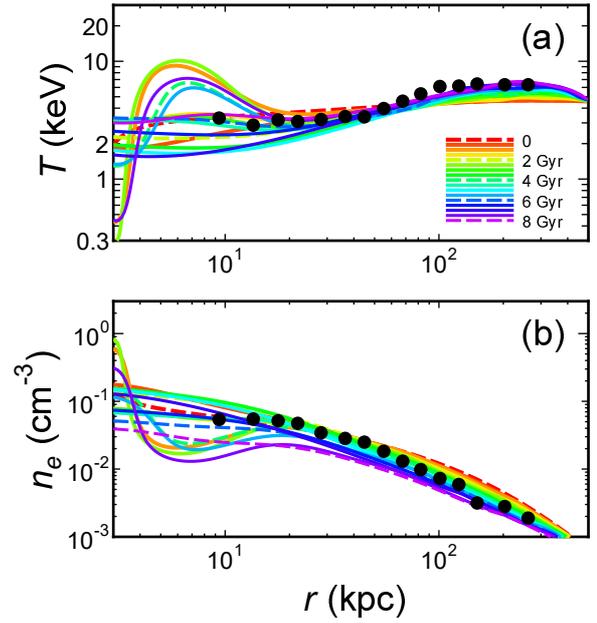} \caption{Same as
 Figure~\ref{fig:Tn_M35} but for model~M33 (smaller eddy
 size).}  \label{fig:Tn_M33}
\end{figure}

\begin{figure}
 \includegraphics[width=0.9\columnwidth]{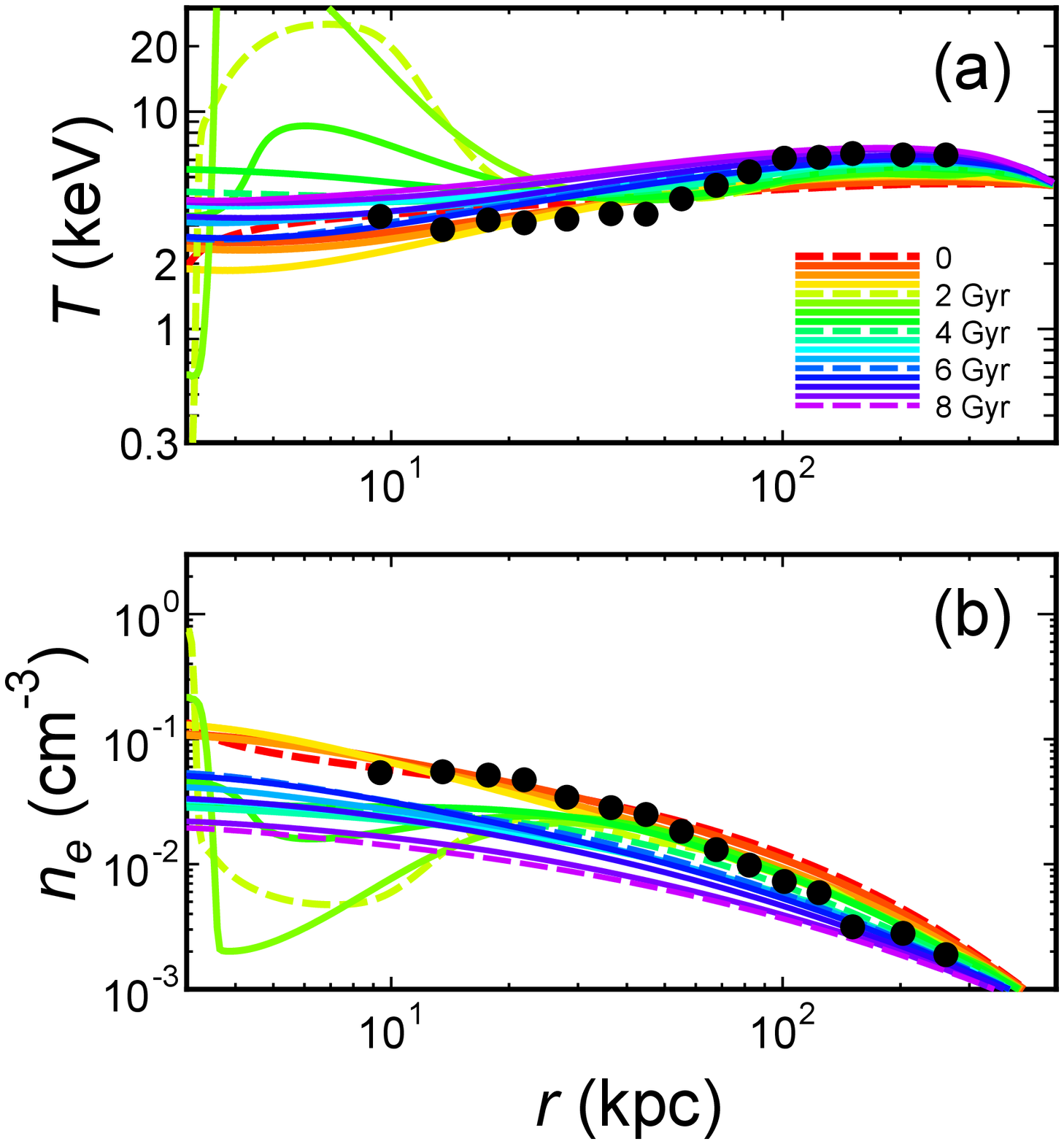} \caption{Same as
 Figure~\ref{fig:Tn_M35} but for model~M35v (time-variational
 turbulence).}  \label{fig:Tn_M35v}
\end{figure}

\begin{figure}
 \includegraphics[width=0.9\columnwidth]{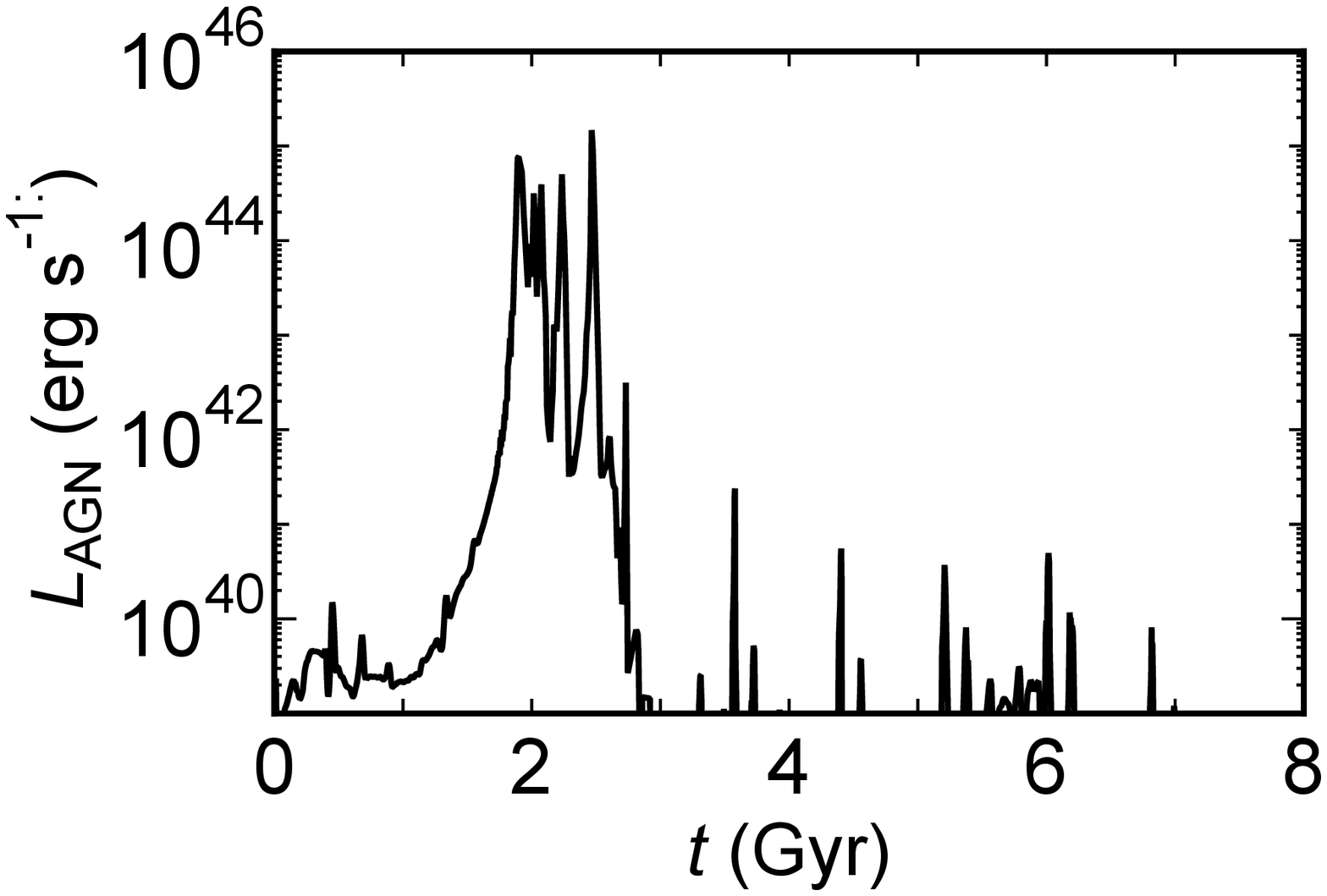} \caption{Same as
 Figure~\ref{fig:LAGN_M35} but for model M35v (time-variational
 turbulence).}  \label{fig:LAGN_M35v}
\end{figure}

\begin{figure}
 \includegraphics[width=0.9\columnwidth]{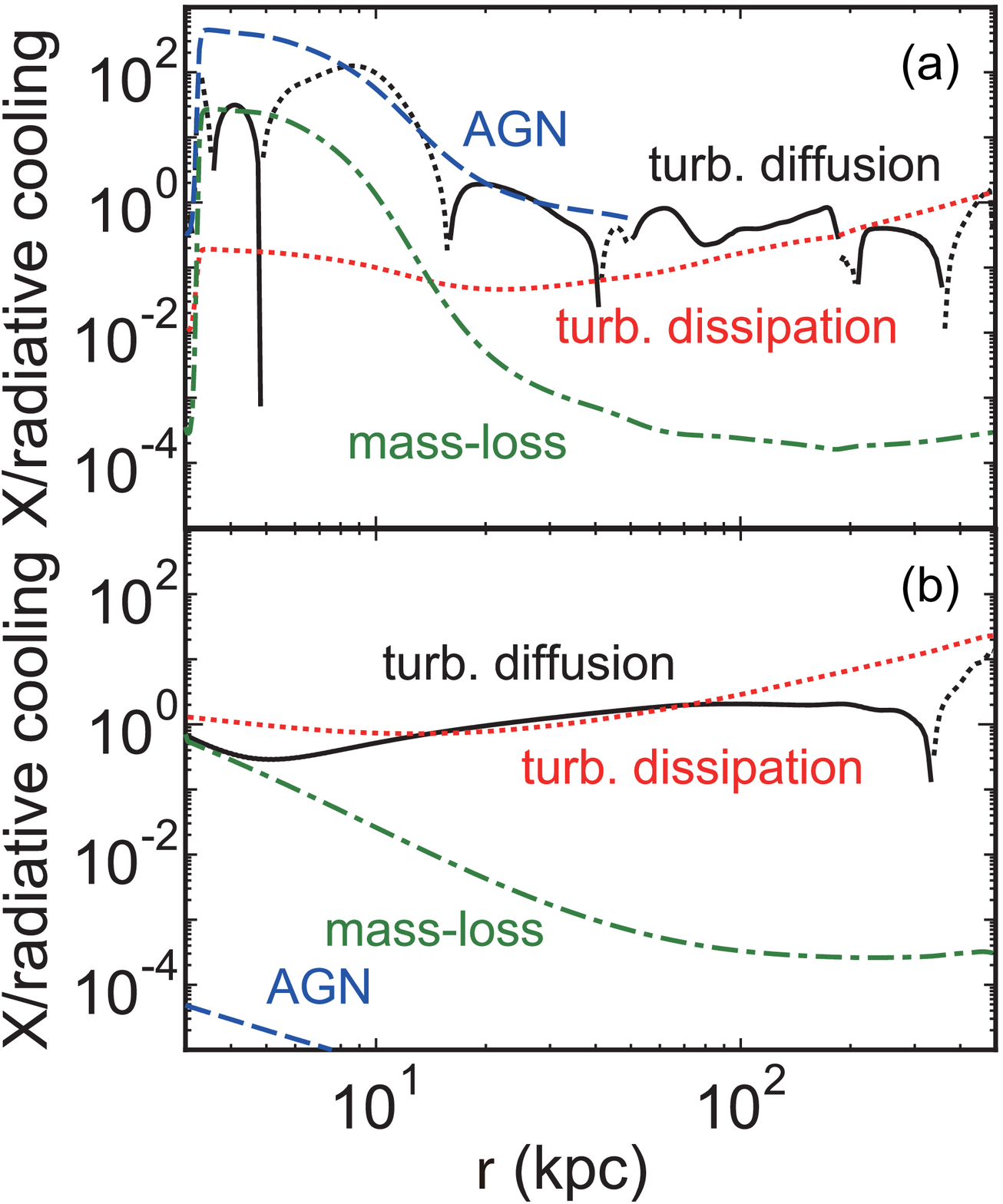} \caption{Same
 as Figure~\ref{fig:balance_M35} but for model M35v (time-variational
 turbulence) at (a) $t=2$~Gyr and
 (b) $t=4$~Gyr.}  \label{fig:balance_M35v}
\end{figure}

\subsection{Dependence on turbulence parameters}
\label{sec:para-dep}

Figure~\ref{fig:Tn_M25} shows the temperature and density profiles for
model~M25. This model gives weaker turbulence ($\alpha_u=0.2$) than the
fiducial model~M35 ($\alpha_u=0.3$). Since the turbulent diffusion is
insufficient, the temperature and density in the central region
fluctuate with a larger amplitude than those for model~M35, and hot and
cool gas are not well mixed. As a result, the temperature at
$r=4$~kpc, which is larger than the inner boundary ($r=3$~kpc), reaches
zero at $t=2.2$~Gyr. Since our code does not include the effect of
drop-out of cold gas from hot gas, the calculation stops at that
point. This shows that a certain level of turbulence is required for
smooth heating to be consistent with the existence of a temperature
floor in observed clusters, although this is not necessarily  the case in some 3D simulations \citep[e.g.][see also discussion in Section~\ref{sec:discuss}]{2016ApJ...829...90Y}. We note that if we adopt stronger AGN
heating $\epsilon=0.1$, the radius where the temperature goes to zero,
moves outward because of strong heating at the cluster centre. Note we extrapolated the cooling function (equation~[\ref{eq:cool}]) to $<10^5$~K. Thus, our calculation at that temperature range is not precise. However, some 3D simulations have suggested that the cold gas should be affected by local thermal instability and should not enter back into hot phase. Therefore, the evolution of the overall ICM is not sensitive to the details of the cooling function \citep[e.g.][]{2012MNRAS.419.3319M}.

The temperature and density profiles for model~M55 ($\alpha_u=0.5$) are
presented in Figure~\ref{fig:Tn_M55}. Since the turbulent heating
dominates radiative cooling, the temperature significantly increases
while the density decreases. Although clusters are growing and their
temperature increases, the rapid evolution in Figure~\ref{fig:Tn_M55}
seems to be inconsistent with actual clusters
\citep[e.g.][]{1998ApJ...503..569E}. Thus, it may be unrealistic that
clusters constantly have turbulence as strong as that in model~M55. The
appropriate level of turbulence should be discussed in the context of
cluster mass accretion rate.  The contributions of turbulent
dissipation and turbulent diffusion to the heating are comparable. While
turbulent dissipation converts energy into heat, turbulent diffusion carries the
energy. Thus, if the thermal energy of the ICM is increased by the
former, the energy carried by the latter also increases. Although the
gas distribution dramatically changes in model~M55, it hardly affects
$V_{\rm circ}$ because the baryon mass fraction of the cluster is only
0.13 as is mentioned above.

Figure~\ref{fig:Tn_M33} shows the results of model~M33 ($\alpha_l=0.3$)
in which the size of eddies are smaller than that in the fiducial model
M35 ($\alpha_l=0.5$). The results of the two models are similar
(Figures~\ref{fig:Tn_M33} and~\ref{fig:Tn_M35}). Thus, turbulent heating
is not sensitive to the eddy size compared with the velocity of the
turbulence ($\alpha_u$).

In real clusters, turbulence in the ICM is expected to vary. For
example, it is likely that turbulence is intensified during cluster
mergers. In order to study the effects of time-variation, we study
model~M35v in which turbulent velocity is given by
\begin{equation}
 u(r) = \alpha_u (1 + 0.5\sin(t/t_p)) V_{\rm cir}(r)\:,
\end{equation}
where $t_p=0.5$~Gyr; other parameters are the same as model M35.
Here, we implicitly assume that the turbulence is 
driven by sloshing of a cool core caused by a minor cluster merger. Thus,
the value of $t_p$ roughly corresponds to a dynamical time-scale of the
central region of a cluster. We note that this study does not focus on
major clusters mergers \citep[e.g.][]{2017MNRAS.464..210V}.
Figures~\ref{fig:Tn_M35v} and~\ref{fig:LAGN_M35v} show that the cluster
is affected by strong AGN bursts at $t\sim 2$~Gyr. The central ICM is
heated (Figure~\ref{fig:balance_M35v}a) and the cool core is almost
destroyed by the bursts. Then, the ICM slowly cools through radiative
cooling and the cool core is reshaped. The AGN does not burst strongly
during this period and the core is in a quasi-steady state
(Figure~\ref{fig:balance_M35v}b). These results suggest that AGN
activities are not necessarily periodic in real clusters as opposed to
model~M35 (Figure~\ref{fig:LAGN_M35}).

\subsection{AGN efficiency and thermal conduction}
\label{sec:cond}

\begin{figure}
 \includegraphics[width=0.9\columnwidth]{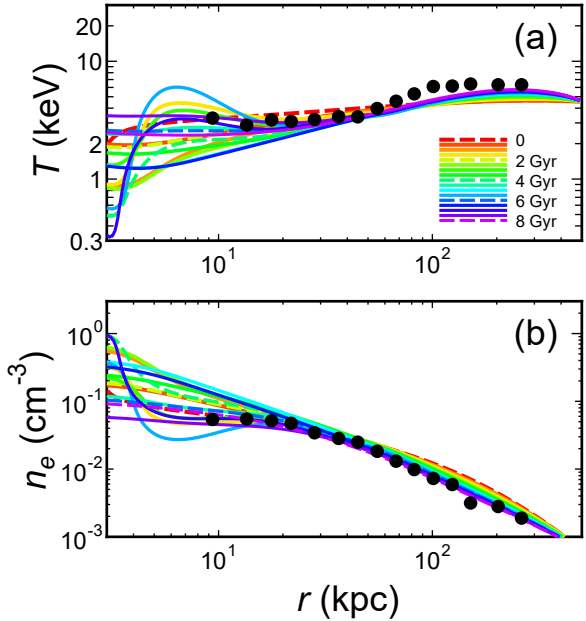} \caption{Same as
 Figure~\ref{fig:Tn_M35} but for model~M35e (larger AGN efficiency). }
 \label{fig:Tn_M35e}
\end{figure}

\begin{figure}
 \includegraphics[width=0.9\columnwidth]{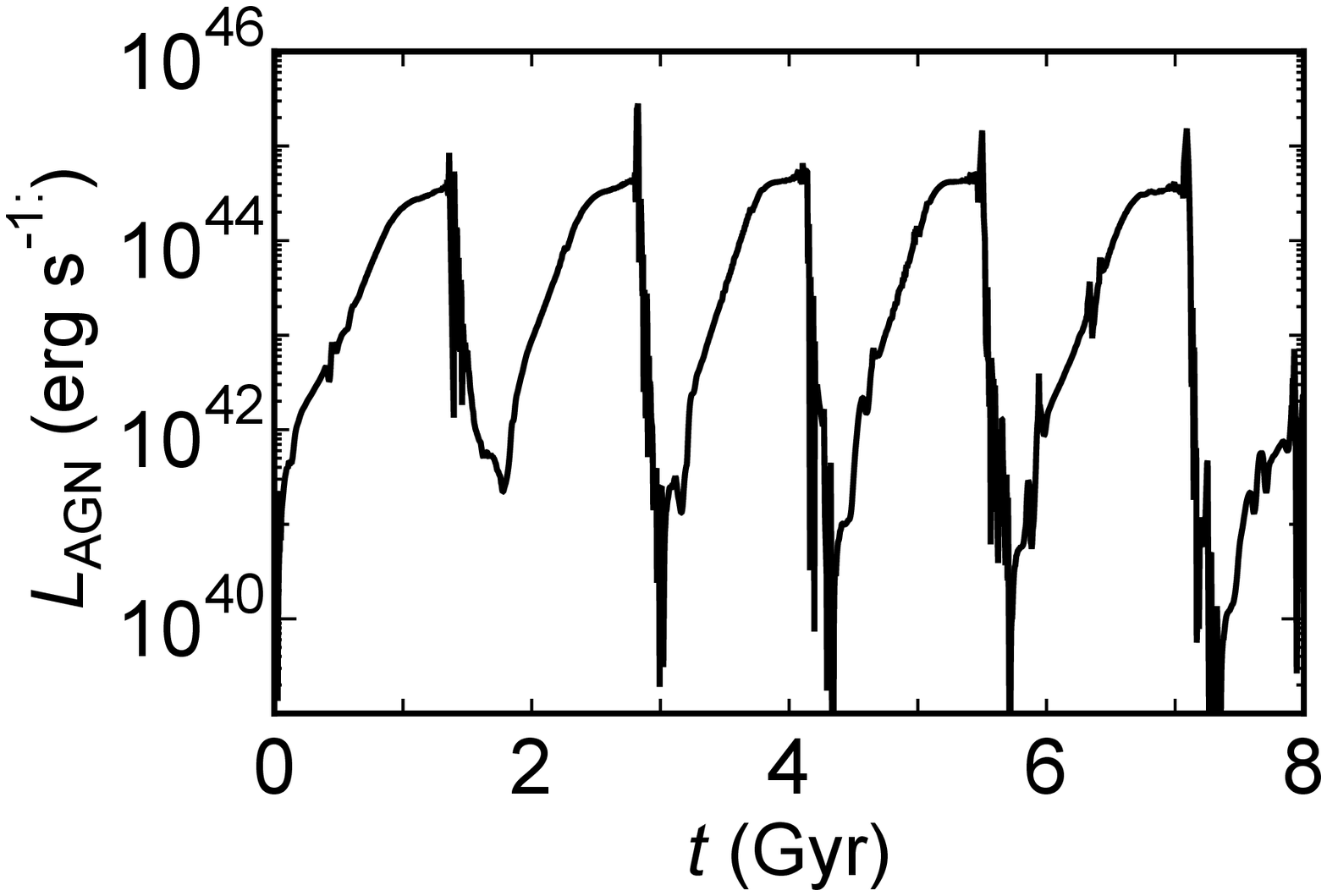} \caption{Same as
 Figure~\ref{fig:LAGN_M35} but for model~M35e (larger AGN efficiency). }  \label{fig:LAGN_M35e}
\end{figure}

\begin{figure}
 \includegraphics[width=0.9\columnwidth]{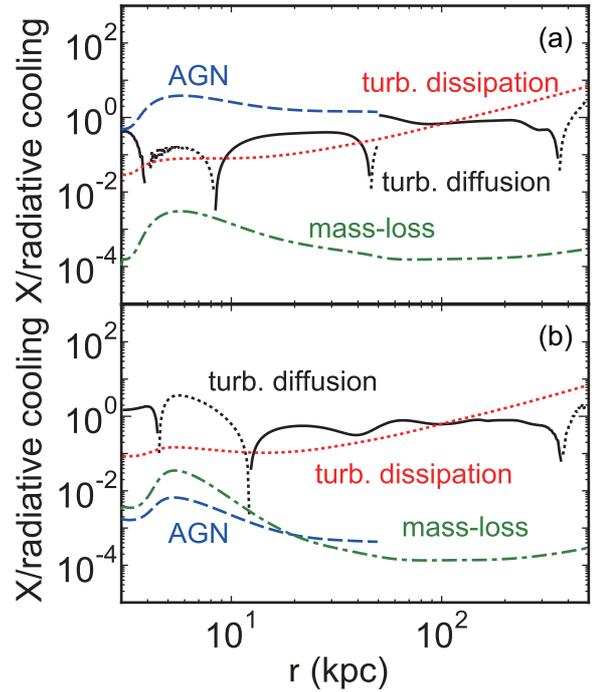} \caption{Same
 as Figure~\ref{fig:balance_M35} but for model M35e (larger AGN efficiency) at (a) $t=4$~Gyr and
 (b) $t=3$~Gyr.}  \label{fig:balance_M35e}
\end{figure}

Figures~\ref{fig:Tn_M35e}--\ref{fig:balance_M35e} show the results for
model~M35e, in which the accretion efficiency ($\epsilon=0.1$) is larger
than the fiducial model M35 ($\epsilon=0.01$). The efficiency
$\epsilon=0.1$ is probably the maximum value in the sense that all
inflow gas is swallowed by the black hole and all the energy generated
by the AGN with the maximum radiative efficiency of an accretion disc \citep[$\sim 10$\%; e.g.][]{2014ARA&A..52..529Y} goes to the ICM. The temperature
and density profiles are similar between the two models
(Figures~\ref{fig:Tn_M35e} and~\ref{fig:Tn_M35}). While the AGN shows
periodic bursts in both models (Figures~\ref{fig:LAGN_M35e}
and~\ref{fig:LAGN_M35}), the activity is less spiky in model~M35e. In
Figure~\ref{fig:balance_M35e}a, we show the energy balance during a
burst phase. The AGN generates energy large enough to counterpart the
radiative cooling with less accretion rate. Thus, the response to
radiative cooling and the subsequent increase of the mass accretion rate
becomes milder. During a quiet phase, the contribution of the AGN is
minor (Figure~\ref{fig:balance_M35e}b). We have also studied the
evolution when $\epsilon=0.001$. In this case, we find that the AGN
activity is too spiky to be smoothed out by turbulence.

\begin{figure}
 \includegraphics[width=0.9\columnwidth]{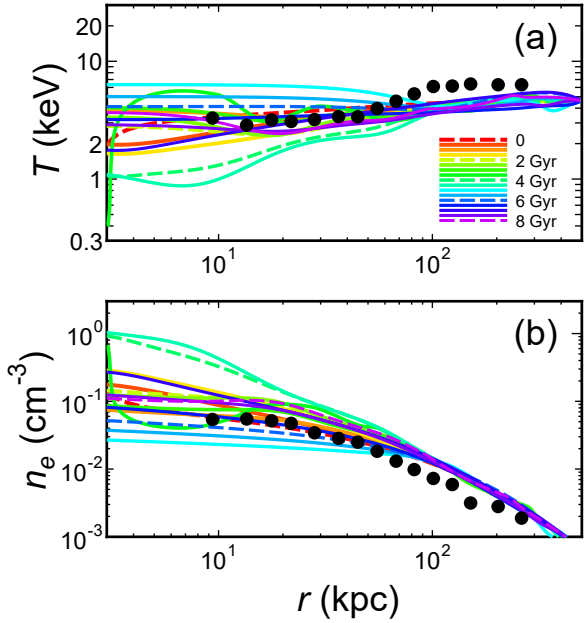} \caption{Same as
 Figure~\ref{fig:Tn_M35} but for model~M25c 
 (turbulence + conduction). } \label{fig:Tn_M25c}
\end{figure}

\begin{figure}
 \includegraphics[width=0.9\columnwidth]{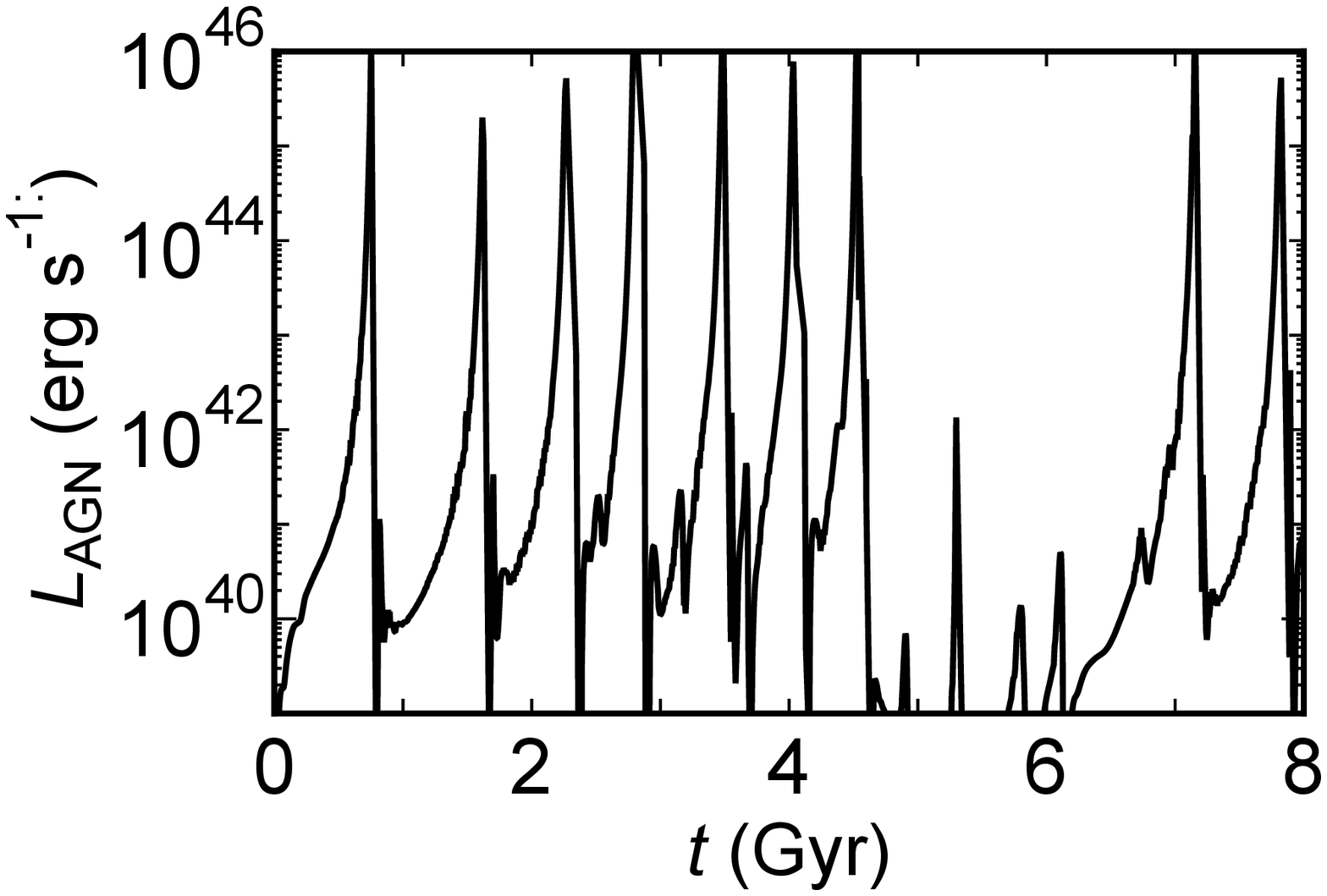} \caption{Same as
 Figure~\ref{fig:LAGN_M35} but for model~M25c 
 (turbulence + conduction).}  \label{fig:LAGN_M25c}
\end{figure}

\begin{figure}
 \includegraphics[width=0.9\columnwidth]{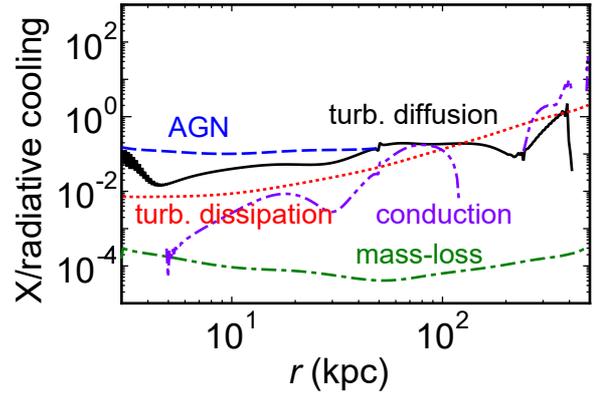} \caption{Same
 as Figure~\ref{fig:balance_M35} but for model M25c 
 (turbulence + conduction) at
 $t=4$~Gyr. Two-dotted-dashed purple line shows thermal conduction
 (second term on the right hand side of equation~[\ref{eq:eg}])}
 \label{fig:balance_M25c}
\end{figure}

Figures~\ref{fig:Tn_M25c}--\ref{fig:balance_M25c} show the results for
model~M25c. In this model, thermal conduction of $f_c=0.2$ (see
equation~[\ref{eq:eg}]) is included. The value of $f_c$ is expected when
magnetic fields are chaotic \citep{2001ApJ...562L.129N}. In model~M25
($f_c=0$), hot and cool gas is not well mixed and the calculation stops
at 2.2~Gyr (Figure~\ref{fig:Tn_M25}). In model~M25c, on the contrary,
thermal conduction assists the mixture and the calculation completes a
period of 8~Gyr (Figures~\ref{fig:Tn_M25c} and~\ref{fig:LAGN_M25c}),
although the fluctuations of temperature and density in the central
region are larger (Figure~\ref{fig:Tn_M25c}).
Figure~\ref{fig:balance_M25c} shows the balance between heating and
cooling at $t=4$~Gyr; thermal conduction significantly contributes as a
heating source at $r\sim 80$~kpc.  We note that if no
turbulence is included and only thermal conduction ($f_c=0.2$) is
considered, cluster temperature at the centre falls to zero, although
the time is delayed compared with the case when both turbulence and
conduction are not included. The conductivity is sensitive to
temperature (equation~[\ref{eq:kappa}]). Since we start the calculation
when the central temperature cools down to 2~keV (see
Section~\ref{sec:result}), 
the conduction alone cannot find or sustain a
steady-state of the ICM.

\subsection{Dependence on cluster mass}
\label{sec:mass}

\begin{figure}
 \includegraphics[width=0.9\columnwidth]{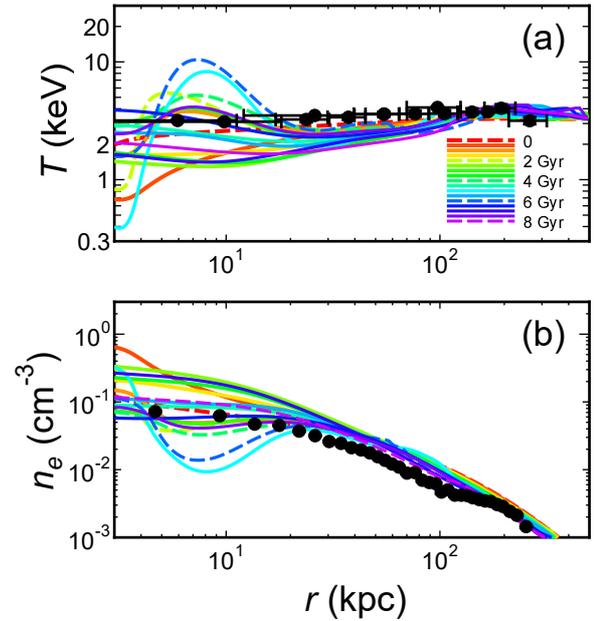} \caption{Same as
 Figure~\ref{fig:Tn_M35} but for model~L35 (low temperature).
 Black dots are observations for the Hydra~A cluster
 \citep{2001ApJ...557..546D}.}  \label{fig:Tn_L35}
\end{figure}

\begin{figure}
 \includegraphics[width=0.9\columnwidth]{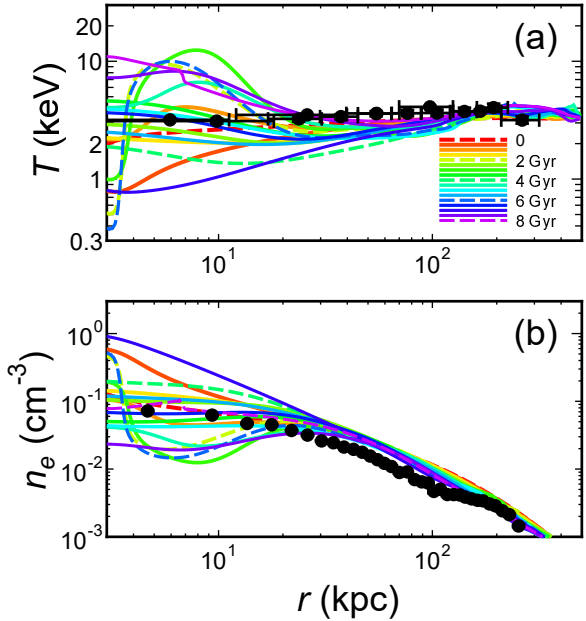} \caption{Same as
 Figure~\ref{fig:Tn_L35} but for model~L35b (low
 temperature + larger $M_{\rm BCG}$ and $M_{\rm BH}$).}
 \label{fig:Tn_L35b}
\end{figure}

In Figure~\ref{fig:Tn_L35}, we present the results for model~L35, in
which the cluster mass ($M_{200}=5.5\times 10^{14}\rm\: M_\odot$) is
smaller than that in model~M35 ($M_{200}=8.5\times 10^{14}\rm\:
M_\odot$). This mass gives $c_{200}=4.2$ and $T_c=3.3$~keV
(Section~\ref{sec:pot}). The gravitational potential of the
cluster is accordingly modified from that in model~M35
(equation~[\ref{eq:gtot}]). The stellar mass-loss rate is also changed
because it depends on $M_{200}$ (Section~\ref{sec:ML}). Other initial
settings (e.g. gas mass fraction) are the same.  As a reference, we
show the observational results for Hydra~A cluster
\citep{2001ApJ...557..546D}, which has a temperature close to
$T_c$. The turbulent velocity profile depends on the
gravitational potential (equation~[\ref{eq:ccir}]) and is shown in the
dotted red line in Figure~\ref{fig:u1D}.  The evolution is similar to
the fiducial model (model M35; Figures~\ref{fig:Tn_M35}). However, this
result depends on our assumption that the AGN heating is less powerful
in less massive clusters ($M_{\rm BCG}\propto M_{200}^{0.4}$ and $M_{\rm
BH}\propto M_{\rm BCG}$)\footnote{Although the efficiency $\eta$
also depends on $M_{\rm BH}$ (equations~[\ref{eq:dotMEdd}] and
[\ref{eq:eta}]), the dependence does not affect the results significantly.}.
Observations have shown that the $M_{\rm BCG}$--$M_{200}$ relation has a
large scatter \citep{2018AstL...44....8K,2019A&A...631A.175E}, and some
less massive clusters have a fairy massive BCG. In order to study this
case, we consider a model in which $M_{\rm BCG}$ and $M_{\rm BH}$ are
the same as model M35, while other parameters are the same as model
L35. We call this model L35b and show the results in
Figure~\ref{fig:Tn_L35b}. Compared with Figure~\ref{fig:Tn_L35}, the
fluctuations of temperature and density are larger, which reflects
stronger AGN bursts. The larger $M_{\rm BH}$ and the
higher gas density associated with the deeper BCG potential cause 
more intensive bursts. If the strong AGN activities are associated with
jets launched by the AGN, large-scale shocks may be created due to their
kinetic power. These shocks may be the ones observed in some less
massive clusters (e.g. MS0735.6$+$7421, Hercules~A, Hydra~A, and
SPT-CLJ0528-5300;
\citealp{2005Natur.433...45M,2005ApJ...625L...9N,2005ApJ...628..629N,2009A&A...495..721S,2019ApJ...887L..17C}).

\begin{figure}
 \includegraphics[width=0.9\columnwidth]{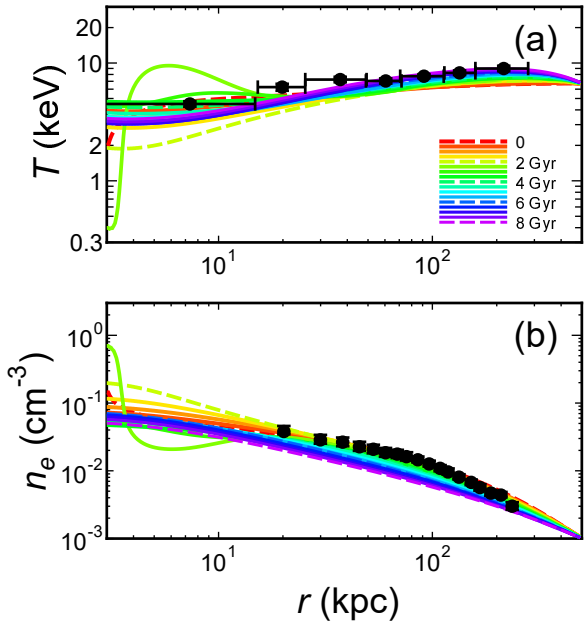} \caption{Same as
 Figure~\ref{fig:Tn_M35} but for model~H35 (high temperature).
 Black dots are observations for the Abell~2029 cluster (north
 direction) \citep{2013ApJ...773..114P}.}  \label{fig:Tn_H35}
\end{figure}

Figure~\ref{fig:Tn_H35} shows the results for model~H35 in which the
cluster mass ($M_{200}= 1.4\times 10^{15}\: M_\odot$, $c_{200}=3.8$ and
$T_c=6.8$~keV) is larger than that in model~M35 ($M_{200}=8.5\times
10^{14}\rm\: M_\odot$). We show the observational results for the
Abell~2029 cluster as a reference \citep{2013ApJ...773..114P}. In this
model, the influence of the AGN bursts is less significant, because the
contribution of the AGN heating is less prominent ($M_{\rm
BH}/M_{200}\propto M_{200}^{-0.6}$) and the turbulent velocity is larger
(Figure~\ref{fig:u1D}); the latter effect is dominant.

Our simulations suggest that clusters should have a large
spread in radial profiles especially for temperatures. This is true
particularly for relatively low-temperature clusters
(Figures~\ref{fig:Tn_L35} and~\ref{fig:Tn_L35b}). Observations appear to
support this prediction (e.g. Figure~3 in
\citealt{2010A&A...517A..92A}).

\section{Discussion}
\label{sec:discuss}

Since clusters of galaxies are still growing, we expect that a certain
level of turbulence always exists in the ICM. We have considered AGN
feedback in the cool core of a cluster when moderate turbulence prevails
in the core, focusing on the global stability of the core. The
results of our numerical simulations show that the AGN activity is
intermittent with occasional bursts. During the quiet phase in general,
radiative cooling of the cool core is nearly balanced with the heating
through turbulent diffusion and turbulence dissipation (sum of both),
and the AGN contribution to the heating is minor. Turbulent diffusion
conveys thermal energy from the outside of the core. However, when the
turbulent velocity is moderate ($\alpha_u\sim 0.3$), the turbulent
heating (diffusion plus dissipation) cannot completely offset the
radiative cooling, although it can prolong the quiet period and reduce
the frequency of the AGN bursts. Similar bursty behaviours of the AGN
have been studied for isolated elliptical galaxies
\citep{2007ApJ...665.1038C,2017ApJ...835...15C}. In this case, the AGN alone can heat and blow away most of the hot interstellar medium, because of the lack of external pressure. On the other hand, the BCG at the centre of a cool core is
affected by external factors such as an inflow from the outside and
turbulence induced by cluster growth. These effects influence the
frequency and duration of the AGN bursts as we have shown in this study.

Our model of AGN heating is very simple (section~\ref{sec:AGN}), and our
results suggest that even such an ``unsophisticated'' model works. That
is, subtle balance between the AGN heating and radiative cooling is not
required, because turbulence smoothes out strong temperature and density
inhomogeniety and a cool core does not need to be in a steady state. In
the vicinity of the AGN ($r\ll 1$~kpc), which cannot be resolved in our
simulations, thermal instabilities may develop and stimulate AGN
activities
\citep{2012MNRAS.419.3319M,2012MNRAS.420.3174S,2012MNRAS.424..728B,2013MNRAS.432.3401G,2014ApJ...780..126G,2015ApJ...808...43M}. While
these activities may increase the AGN heating rate, they can be regard
as local phenomena as long as their contribution is minor compared to
the turbulent heating. Thus, we do not expect tight correlations between
AGN properties and global properties of the host cluster such
as the total mass and the average temperature.

Our simulations show that the velocity of turbulence must be in a
certain range ($\alpha_u\sim 0.3$) in order to avoid violent heating
and/or cooling (section~\ref{sec:para-dep}). If the velocity is too
small, turbulence cannot sufficiently mix hot and cool gas, which leads
to global catastrophic cooling (Figure~\ref{fig:Tn_M25}). If it is too
large, the cluster is strongly heated through turbulent dissipation and
the cluster temperature rises too rapidly
(Figure~\ref{fig:Tn_M55}). This indicates that clusters cannot sustain
turbulence with such a large $\alpha_u$ for a long time considering an
actual mass accretion rate and history of the clusters. However,
$\alpha_u$ may temporally be boosted by a cluster merger. In this case,
the cool core would be destroyed and the cluster turns into a non-cool
core cluster (even though our formulation is developed for cool-core
clusters).

The level of turbulence in clusters will be predicted more precisely by
cosmological numerical simulations with a high resolution. Previous
studies showed that the ratio of turbulent pressure to total pressure in
the central region of clusters is typically an order of $\sim 10$~\%,
although there is a considerable variation among clusters
\citep[e.g.][]{2014ApJ...792...25N,2018MNRAS.481L.120V}. The value is
roughly consistent with our assumption. Observationally, the level will
be measured with high-spectral resolution X-ray missions such as
\textit{X-ray Imaging and Spectroscopy Mission
(XRISM)}\footnote{http://xrism.isas.jaxa.jp/en/} and
\textit{Athena}\footnote{https://www.the-athena-x-ray-observatory.eu/}
in the future. In particular, it could be checked whether turbulent
regions have a flat entropy profile or they correspond to high-entropy
regions that are radially connected to each other if turbulence actually
transports heated gas around the AGN and/or outside the core. Since
there could be strong azimuthal variation, mapping observations of
turbulence are essential. Moreover, AGN activities in clusters with
strong turbulence (generated through cluster growth) tend to be
weak. This could also be observed in the future X-ray missions.

We note that the value of $\alpha_u\sim 0.3$ may change in more
realistic 3D simulations that reproduce complicated gas motions in
clusters. Also our simulations do not explicitly include
turbulence and gas motion 
triggered by jet activities of the central
AGN. If the contribution is significant, the turbulence generated
through cluster growth does not need to be as strong as we
assumed. Thermal conduction can also reduce the requisite level of
turbulence. In order to study the turbulence driven by an AGN, X-ray
observations of turbulence around the AGN will be useful. Moreover, more
realistic models of AGN feedback that include generation of turbulence,
bubbles and shocks should be considered. The heating by AGN jets in
turbulent ICM has been studied with 3D numerical simulations
\citep{2017ApJ...849...54L,2017MNRAS.472.4707B}. Although they individually
investigated the feedback only for one case, they showed
that the pre-existing turbulence associated with cluster growth can
enhance the mixing and advection of AGN feedback energy, which is
consistent with our results. On the other hand, if observations prove
that actual turbulence in clusters is generally not enough to sustain
stable heating, it may mean that thermal conduction is working in
cluster cores (section~\ref{sec:cond}).

In our model, less massive clusters can show violent AGN activities if
the ratio between the BCG mass and the host cluster mass is large
(section~\ref{sec:mass}). This means that the cool cores of those
clusters tend to be destroyed. This tendency may be confirmed with {\it
eROSITA} by observing many clusters and measuring their $M_{\rm
BCG}/M_{200}$ with various methods, while the destruction of cool cores
by cluster mergers also needs to be taken into account. Moreover,
large-scale shocks created through past strong AGN bursts
(e.g. MS0735.6$+$7421; \citealp{2005Natur.433...45M}) would commonly be
observed in those clusters. The ratio $M_{\rm BCG}/M_{200}$ may be
larger at high redshifts because galaxies form earlier than clusters in
a standard hierarchical clustering scenario. Thus, a larger fraction of
cool cores may be destroyed by AGN activities.

Finally, we should discuss caveats on the use of 1D simulations
in this work. Recently, 
 3D simulations of AGN activity in the cluster centre have shown that AGN jets can stably heat the cool
core without the aid of turbulence excited by cluster growth
\citep[e.g.][]{2016ApJ...829...90Y}. These simulations indicate that the
jet activities induce circulation of heated gas, which could be observed
with {\it XRISM} and {\it Athena}. This kind of gas motion including simultaneous inflow and outflow cannot be
reproduced in our 1D simulations.  The circulation may enable stable
heating of the cool core, even if the initial heat injection is
concentrated at the cluster centre. Full 3D simulations show that in
clusters generally $\sim 50$\% of jet energy is communicated to the ICM
during the inflation process
\citep{2013MNRAS.430..174H,2017MNRAS.470.4530W,2017MNRAS.472.4707B,2019MNRAS.490..343B}. In
that sense, we do not intend to conclude that turbulence created through
cluster growth is always necessary for stable heating. Moreover, our
simulations do not capture
non-linear evolution of
local instabilities that affect the global evolution
of the ICM \citep[e.g.][]{2012MNRAS.419.3319M}.

However, despite many attempts to properly simulate the
physics of AGN feedback, the consistency of these simulations with
observations remains questionable. For example, many 3D simulations
reproduce well-extended FR II-like jets instead of short FR I-like jets
(or radio lobes) observed in cluster cores. Many simulations assume that
the jet matter is in a form of thermal gas, while radio and X-ray
observations show that jets are filled with relativistic
particles. Another issue is the energy injection into bubbles/jets. In
simulations, the injection process is often very fast and rather
violent, while X-ray observations are consistent with a gentle scenario,
during which most of the energy from SMBH is injected into bubbles
instead of shocks and shock-heated gas. Moreover, the presence of weak
magnetic fields has been confirmed with the radio observations, however,
these fields are often neglected in the feedback simulations. Since it
is very difficult to properly take all the physics into account (and
often unclear what to include), 1D "toy models" focused on specific
aspects of the gas heating and on comprehensive survey can give
interesting new insights into the problem. In this work, in particular,
we show that turbulence created by the mergers and other processes
relevant to the growth of clusters can contribute to the heating of
clusters cores. Our simulations do not exclude a possibility that jets
alone can stably heat the gas. However, the contribution from the
turbulence triggered by structure formation is likely energetically
important.

\section{Conclusions}

We have studied a new class of time-dependent cool core models in which
radiative cooling is offset by a combination of central AGN heating and
moderate turbulence excited through cluster growth.  We
investigated the global stability of the core for a wide range of
cluster and turbulence parameters. We found that a cool core does not
achieve a steady state and the AGN shows intermittent activities.  The
core is in a quasi-equilibrium state for most of the time because the
heating through turbulent diffusion and dissipation are nearly balanced
with radiative cooling. The contribution of the AGN heating is minor
during this phase.  The balance between cooling and heating is
eventually lost because of slight dominance of the cooling.  The mass
accretion rate toward the central black hole increases and finally the
AGN bursts. As a result, the core is almost instantaneously
heated. Since the already-existing turbulence mixes the heated gas with
surrounding gas, the core does not become unstable and recovers the quasi-equilibrium state. The
AGN bursts can be stronger in lower-temperature clusters if the ratio of
the BCG mass to the cluster mass is large. Future X-ray missions such as
{\it XRISM} and {\it Athena} will be able to test our predictions.

Our study is based on 1D simulation and our model of AGN feedback is
rather simple, For more quantitative discussion, 3D cosmological
simulations and more realistic AGN feedback models would be desirable.

\section*{Acknowledgements}

We would like to thank the anonymous referee for a constructive
report. This work was supported by MEXT KAKENHI Nos. 18K03647 (Y.F.). IZ is partially supported by a Clare Boothe Luce Professorship from the Henry Luce Foundation.




\bibliographystyle{mnras}
\bibliography{list} 




%
%
%


\bsp	
\label{lastpage}
\end{document}